\expandafter\ifx\csname LaTeX\endcsname\relax
      \let\maybe\relax
\else \immediate\write0{}
      \message{You need to run TeX for this, not LaTeX}
      \immediate\write0{}
      \makeatletter\let\maybe\@@end
\fi
\maybe

\magnification=\magstephalf

\hsize=5.25truein
\vsize=8.3truein
\hoffset=0.37truein

\newdimen\frontindent \frontindent=.45truein
\newdimen\theparindent \theparindent=20pt


\let\em=\it

\font\tencsc=cmcsc10
\font\twelvebf=cmbx10 scaled 1200
\font\bmit=cmmib10  \font\twelvebmit=cmmib10 scaled 1200
\font\sixrm=cmr6 \font\sixi=cmmi6 \font\sixit=cmti8 at 6pt

\font\eightrm=cmr8  \let\smallrm=\eightrm
\font\eighti=cmmi8  \let\smalli=\eighti
\skewchar\eighti='177
\font\eightsy=cmsy8
\skewchar\eightsy='60
\font\eightit=cmti8
\font\eightsl=cmsl8
\font\eightbf=cmbx8
\font\eighttt=cmtt8
\def\eightpoint{\textfont0=\eightrm \scriptfont0=\fiverm 
                \def\rm{\fam0\eightrm}\relax
                \textfont1=\eighti \scriptfont1=\fivei 
                \def\mit{\fam1}\def\oldstyle{\fam1\eighti}\relax
                \textfont2=\eightsy \scriptfont2=\fivesy 
                \def\cal{\fam2}\relax
                \textfont3=\tenex \scriptfont3=\tenex 
                \def\it{\fam\itfam\eightit}\let\em=\it
                \textfont\itfam=\eightit
                \def\sl{\fam\slfam\eightsl}\relax
                \textfont\slfam=\eightsl
                \def\bf{\fam\bffam\eightbf}\relax
                \textfont\bffam=\eightbf \scriptfont\bffam=\fivebf
                \def\tt{\fam\ttfam\eighttt}\relax
                \textfont\ttfam=\eighttt
                \setbox\strutbox=\hbox{\vrule
                     height7pt depth2pt width0pt}\baselineskip=9pt
                \let\smallrm=\sixrm \let\smalli=\sixi
                \rm}


\catcode`@=11
 
\def\vfootnote#1{\insert\footins\bgroup\eightpoint
     \interlinepenalty=\interfootnotelinepenalty
     \splittopskip=\ht\strutbox \splitmaxdepth=\dp\strutbox
     \floatingpenalty=20000
     \leftskip=0pt \rightskip=0pt \parskip=1pt \spaceskip=0pt \xspaceskip=0pt
     \everydisplay={}
     \smallskip\textindent{#1}\footstrut\futurelet\next\fo@t}
 
\newcount\notenum

\def\note{\global\advance\notenum by 1
    \edef\n@tenum{$^{\the\notenum}$}\let\@sf=\empty
    \ifhmode\edef\@sf{\spacefactor=\the\spacefactor}\/\fi
    \n@tenum\@sf\vfootnote{\n@tenum}}


\tabskip1em

\newtoks\pream \pream={#\strut}
\newtoks\lpream \lpream={&#\hfil}
\newtoks\rpream \rpream={&\hfil#}
\newtoks\cpream \cpream={&\hfil#\hfil}
\newtoks\mpream \mpream={&&\hfil#\hfil}

\newcount\ncol \def\ncolp{\advance\ncol by 1}
\def\atalias#1{
    \ifx#1l\edef\xpream{\pream={\the\pream\the\lpream}}\xpream\ncolp\fi
    \ifx#1r\edef\xpream{\pream={\the\pream\the\rpream}}\xpream\ncolp\fi
    \ifx#1c\edef\xpream{\pream={\the\pream\the\cpream}}\xpream\ncolp\fi}
\catcode`@=\active

\def\taborl#1{\omit\unskip#1\hfil}
\def\taborc#1{\omit\hfil#1\hfil}
\def\taborr#1{\omit\hfil#1}
\def\multicol#1{\multispan#1\let\omit\relax}

\def\table#1\par{\midinsert\offinterlineskip\everydisplay{}
    \let@\atalias \let\l\taborl \let\r\taborr \let\c\taborc
    \def\space{\noalign{\vskip2pt}}
    \def\tablerule{\omit&\multispan{\the\ncol}\hrulefill\cr}
    \def\onerule{\space\space\tablerule\space\space}
    \def\tworules{\space\space\tablerule\space\tablerule\space\space}
    \def\annot##1\\{&\multispan{\the\ncol}##1\hfil\cr}
    \def\\{\let\\=\cr
           \edef\xpream{\pream={\the\pream\the\mpream}}\xpream
           \edef\starthalign{$$\vbox\bgroup\halign\bgroup\the\pream\cr}
           \starthalign
           \annot\hfil\tencsc Table #1\\ \noalign{\medskip}}
    \let\par\endtable}

\edef\endtable{\noalign{\vskip-\bigskipamount}\egroup\egroup$$\endinsert}

\let\plainmidinsert=\midinsert
\def\eightpttable{\def\midinsert{\let\midinsert=\plainmidinsert
    \plainmidinsert\eightpoint\tabskip 1em}\table}



\newif\iftitlepage

\def\raggedright{\rightskip 0pt plus .2\hsize\relax}

\let\caret=^ \catcode`\^=13 \def^#1{\ifmmode\caret{#1}\else$\caret{#1}$\fi}

\def\title#1\par{\vfill\supereject\begingroup
                 \global\titlepagetrue
                 \leftskip=\frontindent\parindent=0pt\parskip=0pt
                 \frenchspacing \eqnum=0
                 \gdef\runningtitle{#1}
                 \null\vskip-22.5pt\copy\volbox\vskip18pt
                 {\titlestyle#1\bigskip}}
\def\titlestyle{\raggedright\bf\twelvebf\textfont1=\twelvebmit
                \let\smallrm=\tenbf \let\smalli=\bmit
                \baselineskip=1.2\baselineskip}
\def\shorttitle#1\par{\gdef\runningtitle{#1}}
\def\author#1\par{{\raggedright#1\medskip}}

\def\shortauthor#1\par{\gdef\runningauthors{#1}}

\def\affil#1\par{{\raggedright\it#1\smallskip}}
\def@#1{\ifhmode\qquad\fi\leavevmode\llap{^{#1}}\ignorespaces}
\def\abstract{\smallskip\medskip{\bf Abstract: }}

\def\maybebreak#1{\vskip0pt plus #1\hsize \penalty-500
                  \vskip0pt plus -#1\hsize}

\def\maintextmode{\leftskip=0pt\parindent=\theparindent
                  \parskip=\smallskipamount\nonfrenchspacing}

\def\maintext#1\par{\bigskip\medskip\maintextmode\noindent}

%

\def\references\par{\bigskip\maybebreak{.1}\parindent=0pt
    \everypar{\hangindent\theparindent\hangafter1}
    \leftline{\bf References}\smallskip}

\def\appendix#1\par{\bigskip\maybebreak{.1}\maintextmode
    \advance\secnum by 1 \bigskip\maybebreak{.1}
    \leftline{\bf Appendix: #1}\smallskip\noindent}

\def\acknowl{\medskip\noindent}

\def\bye{\endgroup\vfill\supereject\end}


\newbox\volbox
\setbox\volbox=\vbox{\hsize=.5\hsize \raggedright
       \sixit\baselineskip=7.2pt \noindent
       Nonlinear Dynamics in Astronomy and Physics,
       A Workshop Dedicated to the Memory of Professor
       Henry E. Kandrup.
       To appear in Annals of the New York Academy of Sciences,
       Eds.\ J.R.~Buchler, S. T. Gottesman, M. E. Mahon}


\input epsf

\def\figureps[#1,#2]#3.{\midinsert\parindent=0pt\eightpoint
    \vbox{\epsfxsize=#2\centerline{\epsfbox{#1}}}
    \def\par{\endgraf\endinsert}{\bf Figure#3.}}

\def\figuretwops[#1,#2,#3]#4.{\midinsert\parindent=0pt\eightpoint
    \vbox{\centerline{\epsfxsize=#3\epsfbox{#1}\hfil
                      \epsfxsize=#3\epsfbox{#2}}}
     \def\par{\endgraf\endinsert}{\bf Figure#4.}}

\def\figurespace[#1]#2.{\midinsert\parindent=0pt\eightpoint
    \vbox to #1 {\vfil\centerline{\tenit Stick Figure#2 here!}\vfil}
    \def\par{\endgraf\endinsert}{\bf Figure#2.}}


\headline={\iftitlepage\hfil\else
              \ifodd\pageno\hfil\tensl\runningtitle
                    \kern1pc\tenbf\folio
               \else\tenbf\folio\kern1pc
                    \tensl\runningauthors\hfil\fi
           \fi}
\footline{\iftitlepage\tenbf\hfil\folio\hfil\else\hfil\fi}
\output={\plainoutput\global\titlepagefalse}


\newcount\eqnum
\everydisplay{\puteqnum}  
\def\puteqnum#1$${#1\global\advance\eqnum by 1\eqno(\the\eqnum)$$}
\def\namethiseqn#1{\xdef#1{\the\eqnum}}

 
\newcount\mpageno
\mpageno=\pageno  \advance\mpageno by 1000
 
\def\advancepageno{\global\advance\pageno by 1
                   \global\advance\mpageno by 1 }

\openout15=inx
\def\index#1{\write15{{#1}{\the\mpageno}}\ignorespaces}


\def\LaTeX{{\rm L\kern-.36em\raise.3ex\hbox{\tencsc a}\kern-.15em
    T\kern-.1667em\lower.7ex\hbox{E}\kern-.125emX}}

\def\[#1]{\raise.2ex\hbox{[}#1\raise.2ex\hbox{]}}

\def\witchbox#1#2#3{\hbox{$\mathchar"#1#2#3$}}
\def\leqsim{\mathrel{\rlap{\lower3pt\witchbox218}\raise2pt\witchbox13C}}
\def\geqsim{\mathrel{\rlap{\lower3pt\witchbox218}\raise2pt\witchbox13E}}

\def\<#1>{\langle#1\rangle}


{\obeyspaces\gdef {\ }}

\catcode`@=12 \let\@=@ \catcode`@=13
\def\+{\catcode`\\=12\catcode`\$=12\catcode`\&=12\catcode`\#=12%
       \catcode`\^=12\catcode`\_=12\catcode`\~=12\catcode`\%=12%
       \catcode`\@=0\tt}
\def\({\endgraf\bgroup\let\par=\endgraf\parskip=0pt\vskip3pt
       \eightpoint \def\/{{\eightpoint$\langle$Blank line$\rangle$}}
       \catcode`\{=12\catcode`\}=12\+\obeylines\obeyspaces}
\def\){\vskip1pt\egroup\vskip-\parskip\noindent\ignorespaces}


\def\ltsima{$\; \buildrel < \over \sim \;$}
\def\simlt{\lower.5ex\hbox{\ltsima}}
\def\gtsima{$\; \buildrel > \over \sim \;$}
\def\simgt{\lower.5ex\hbox{\gtsima}}

\title HENRY KANDRUP'S IDEAS ABOUT RELAXATION OF STELLAR SYTEMS

\shorttitle Henry Kandrup

\author David Merritt

\shortauthor David Merritt

\affil Rochester Institute of Technology

\bigskip

\abstract
Henry Kandrup wrote prolifically on the problem of relaxation of
stellar systems. 
His picture of relaxation was significantly more
refined than the standard description in terms of phase mixing
and violent relaxation.
In this article, I summarize Henry's work in this and related areas.

\bigskip\maintextmode

Henry Kandrup was a leading figure in galactic dynamics,
distinguishing himself both as an original thinker
and as an educator.
His published contributions were prolific and wide-ranging.
While I never formally collaborated with Henry,
we often discussed dynamics during my visits
to the University of Florida or at scientific meetings,
and both of us were conscientious readers of the other's
papers.
Henry's ideas often influenced my own work, and I believe
that I can see evidence of the reverse influence 
in some of Henry's published papers.

Here I will focus on a topic that occupied Henry's attention
throughout his career: dynamical chaos and its 
connection with relaxation and equilibrium of stellar systems.
Henry first addressed this topic in his PhD thesis,
``Stochastic Processes in Stellar Dynamics'' [1], and 
returned to it in one of his last papers,
``Chaos and Chaotic Phase Mixing in Galaxy Evolution 
and Charged Particle Beams'' [2].

Henry's unique contribution was to associate relaxation
in stellar systems with the exponential instability of orbits.
Prior to Henry's work, discussions in the astronomical
literature of dynamical relaxation 
rarely addressed the mixing properties of the flow,
and discussions of stochasticity rarely drew conclusions
about relaxation.
Henry argued that there was a fundamental connection
between the two.
He pointed out that the trajectories of stars even
in ``collisionless'' systems like galaxies could often
mimic the exponentially unstable trajectories of 
molecules in a collisional fluid, due either to
non-integrability of the steady-state potential,
or to time variations in the potential associated
with external perturbations or departures from
a steady state.
Henry demonstrated that this chaos or near-chaos could be very effective 
at inducing evolution to a steady state, in much the same way
that collisions between molecules in a gas erase
memory of the initial conditions.
In this way, Henry established a new and important
link between dynamical chaos, statistical mechanics, 
and the structure and evolution of stellar systems.

Many stellar and galactic systems are smooth and 
symmetric in appearance.
This is surprising at first sight, since the time required
for gravitational encounters to ``smear out'' trajectories
is very long in such low-density systems,
often much longer than the age of the universe.
One of the first to note this puzzle was Fritz Zwicky [3].
Zwicky was so impressed by the regular appearance
of the Coma galaxy cluster that he argued against Hubble's
expanding-universe model -- on the grounds that
the age of the universe in Hubble's model was too
short for encounters between galaxies to
remove irregularities in their spatial distribution.

At the time that Henry began his career, the smooth
appearance of galaxies and galaxy clusters
was generally understood to
be due to a combination of phase mixing and
``violent relaxation''.
Phase mixing is the gradual shearing of points in a fixed, 
integrable potential;
after many orbital periods, phase mixing results
in a coarse-grained density that is constant with respect
to angle over the invariant torus.
``Violent relaxation'' was defined as the more
extreme redistribution of particles 
that occurs when the gravitational potential is
rapidly varying.
King [4], H\'enon [5], Lynden-Bell [6]
and others realized that relaxation under conditions
of a rapidly varying potential
might be very efficient, and the last-named author identified
the relaxation rate directly with the rate of change
of the potential.
Support for this hypothesis was seen in numerical
simulations of the collapse of a cold cloud of stars,
where a nearly steady state is reached after just a few
crossing times.

Henry and I often discussed violent relaxation during my
visits to the University of Florida.
During one of these conversations,
Henry told me that he had been very impressed by a letter written
by Richard Miller to Donald Lynden-Bell on July 21, 1966.
In the letter, Miller argued that the identification of relaxation
with changes in the potential was problematic:

{\narrower\smallskip\noindent
A counter example is furnished by a potential
that depends only upon the time.
Consider a stellar dynamical system  described by
a Hamiltonian $H_0$, and another described by $H=H_0+V(t)$
where $V$ depends only upon the time. 
All motions in the
two systems are identical; no relaxation is induced by $V(t)$, 
contrary to the assertion of your equation 1.\smallskip}

\noindent
Miller went on to note:

{\narrower\smallskip\noindent
I think there is a germ of an idea in your assertion
[that relaxation can be identified with changes in the potential],
but it wants more complete working-out. Essentially, I think
that the kind of term that might replace $\langle\dot\Phi^2\rangle$
in your equations is $\langle(G\rho)^{-1/2}\dot\Phi(a\cdot\nabla\Phi)\rangle$
or some rather fancy term of that character -- displaying
both time and space derivatives, but averaged over the cluster
and measured in time units characteristic of the cluster. ($a$
is some length characterizing the cluster.)\smallskip}

\noindent
Henry told me that he shared Miller's reservations
about equating relaxation with $\dot\Phi$.
As he later put it [7]:

{\narrower\smallskip\noindent
More pragmatically, one infers from $N$-body simulations
that a strongly convulsing mean field potential is not
necessary. One observes a comparably efficient approach
towards a meta-equilibrium on a time scale $\sim t_{cr}$
[the crossing time] both for ``violent'' evolution,
where $\Phi$ exhibits huge changes on a very short time
scale, and for ``nonviolent'' evolution, where $\Phi$
exhibits only relatively small changes. Nonviolent
relaxation can  be just as efficient as violent relaxation.
\smallskip}

\noindent In this passage, 
Henry is referring to what he elsewhere 
called ``chaotic mixing'': the exponential spreading
of an ensemble of initially localized, stochastic trajectories.
He continued:

{\narrower\smallskip\noindent
This means that phase mixing can proceed {\it much}
more efficiently for chaotic flows than for regular flows,
where any approach towards a (near-) equilibrum typically
proceeds as a power law in time.
Chaotic flows should relax much more efficiently than do
regular flows.
It would thus seem that the phase mixing of chaotic flows
$\ldots$ could
serve to provide an explanation of why various systems in nature
seem to approach an equilibrium or near-equilibrium as fast as
they do.
In particular, chaotic mixing could  help explain the remarkable
efficacy of violent relaxation: Why do galaxies look ``so
relaxed'' when the nominal relaxation time $t_R$ is typically
much longer than $t_H$, the age of the Universe?
\smallskip}

\noindent
Rather than identify relaxation with either phase mixing 
in a fixed potential, or ``violent relaxation'' in a time-varying
potential,
Henry proposed that the proper distinction was between 
phase mixing and chaotic mixing,
and that the time dependence of the potential was secondary.
The efficient relaxation observed in simulations of
collapse was due, he argued, to the more chaotic
nature of the phase-space flow when the potential
was rapidly varying, and not simply to the redistribution
of energies that takes place when the potential has a 
time-dependent component.

While the existence of stochastic orbits in galactic 
potentials had been appreciated since the work of 
H\'enon, Contopoulos, Miller and others in the 1960s, 
Henry was one of the first to ask what the {\it consequences}
of the chaos might be for the evolution of an ensemble of orbits
toward a statistical steady state.
Henry began a systematic investigation of this question
by looking at the effects 
of chaos in time-independent potentials; by definition, 
``violent relaxation'' can not occur if the potential is fixed.
In a series of papers with M. E. Mahon and other collaborators [8-11],
Henry investigated the relation between stochasticity of
individual trajectories -- as measured, for instance, by Liapunov
exponents -- and the rate at which an initially compact
{\it ensemble} of stars evolves toward a uniform distribution
over the accessible phase space.
These papers showed that the coarse-grained distribution function
typically exhibits an exponential approach toward
equilibrium at a rate that correlates well with the mean
Liapunov exponent for the ensemble.
When evolved for much longer times, the phase space density
 slowly changes as orbits diffuse into regions that, 
although accessible, are avoided over the shorter time interval.
Henry coined the term ``near-invariant distribution''
to describe the end-point of chaotic mixing of an isoenergetic
ensemble of points.

In another paper [12], Henry compared the efficiency of 
phase mixing and chaotic mixing.
He noted that -- for initially very localized ensembles -- the 
two processes occur at very different rates: chaotic mixing 
takes place on the Liapunov, or exponential divergence, time scale 
while the phase mixing rate falls to zero.
But phase mixing of a group of points with a finite extent can 
be much more rapid.
Furthermore the mixing rate of chaotic ensembles typically 
falls below the Liapunov rate once the trajectories separate; 
this is especially true for those stochastic orbits that are confined 
over long periods of time to restricted parts of phase space.
The effective rates of phase mixing and chaotic mixing might 
therefore be comparable in real galaxies.
Henry noted also that chaotic mixing in 3D potentials can occur 
at substantially different rates in different directions.

Henry recognized that the existence of invariant or near-invariant
distributions, in regions of phase space that were not characterized
by three isolating integrals of motion, implied the existence of
a new class of equilibrium or near-equilibrium states for galaxies.
The classical Jeans theorem requires
that the phase-space density of a stationary stellar system
be expressed solely in terms of the isolating integrals in that 
potential.
Henry pointed out that there existed a far larger class of 
systems that could be in a stationary state.
In the abstract of ``Invariant Distributions and Collisionless
Equilibria'' [13], he wrote:

{\narrower\smallskip\noindent
This paper discusses the possibility of constructing time-independent
solutions to the collisionless Boltzmann equation which depend
on quantities other than global isolating integrals such as
energy and angular momentum.
The key point is that, at least in principle, a self-consistent
equilibrium can be constructed from {\it any} set of time-independent
phase-space building blocks which, when combined, generate the mass
distribution associated with an assumed time-independent potential.
\smallskip}

\noindent While noting that strictly ``time-independent'' 
phase-space building blocks were mathematical idealizations, 
Henry pointed out that his ``near-invariant distributions'' were 
effectively time independent, and could in principle be used
as building blocks in the construction of stationary galaxies.
Indeed he argued that chaotic orbits were in a sense more
natural components of steady-state galaxies than regular orbits, 
since an efficient mechanism (chaotic mixing) exists that can
convert a generic distribution of points in chaotic phase
space into a time-independent one.
By contrast, an ensemble of points on a regular torus
does not evolve toward a coarse-grained steady state:
it simply translates, unchanged, around the torus.
Jeans's theorem {\it postulates} a uniform distribution 
over each torus but says nothing about how this
unlikely distribution is to be achieved.

Henry's insight constituted the most significant updating of Jeans's 
theorem since at least the 1960s, when various authors pointed out
the distinction between isolating and non-isolating integrals.
I propose that a ``generalized Jeans theorem'' be attributed to
Henry:

{\narrower\smallskip\noindent
{\bf Generalized Jeans Theorem:} The phase-space density of a
stationary stellar system is constant within every connected
region.\smallskip}

\noindent
A ``connected region'' is one that can not be decomposed into
two finite regions such that all trajectories remain for
all time in either one or the other.
Invariant tori are such regions, but so are the more complex
parts of phase space associated with stochastic orbits.

As Henry once pointed out to me (with some amusement), 
people have actually been invoking the
generalized version of Jeans's theorem for years without realizing it!
A textbook example of a system satisfying the classical
Jeans theorem (and one that was discussed by Jeans himself [14])
is an axisymmetric galaxy in which $f$ is
a function of the two classical integrals of motion,
the energy $E$ and the angular momentum $J_z$.
Writing $f=f(E,J_z)$ implies that the phase space density is
constant on hypersurfaces of constant $E$ and $J_z$.
But not all orbits in axisymmetric potentials are characterized
by a third isolating integral, hence parts of these hypersurfaces
are associated with chaotic trajectories.
The two-integral approach to axisymmetric modelling -- which
assigns a constant density to these regions -- thus
depends on the generalized form of Jeans's theorem for its
justification.
Henry went on to note [13] that one could
in principle construct novel axisymmetric models,
in which the surfaces of constant $E$ and $J_z$ are {\it not}
sampled uniformly;  for instance, one could  exclude all
the chaotic orbits, or assign different densities to 
different chaotic regions on the same ($E,J_z$) hypersurface.
As far as I know, no one has yet attempted to construct
models of this form, although it would be relatively 
straightforward to do so via orbital superposition.
Nevertheless, it has become clear in the last few
years that chaotic orbits can be major components of
steady-state galaxies, demonstrating that Henry's generalized
theorem is potentially very significant for our understanding
galactic structure.
For instance, one recent study [15]
found that 50\% or more of the mass in steady-state
triaxial nuclei could be associated with chaotic orbits.

Smooth potentials are idealizations of real galaxies.
As seen by a single star, any lumpiness or distortions in
the stellar density would add 
small-amplitude perturbing forces to the mean field.
Such perturbations would not be expected to have much consequence 
for either strongly chaotic or precisely regular orbits, but 
Henry realized that they might have an appreciable effect on 
the evolution of weakly stochastic 
orbits, by scattering trajectories away from a trapped region into a 
region where the mixing is more rapid.
Henry, S. Habib and M. E. Mahon [16-18] investigated this
idea, adding random noise to otherwise smooth potentials
and observing the effects on the mixing rate.
They found that even very weak noise, with 
a characteristic time scale $|{{1\over v}{\delta v\over\delta 
t}}|^{-1}$ of order $10^6$ crossing times, could 
induce substantial changes in the motion of trapped stochastic
orbits over just $\sim 10^2$ orbital periods.

In a strongly time-dependent potential, chaos should be
even more prevalent, if only because time-dependent potentials
lack at least one isolating integral of the motion
(the energy) that is always present in stationary potentials.
In three studies [19-21], Henry and collaborators considered the
effects of two sorts of strongly time-dependent perturbations
on the structure of orbits in 2D and 3D potentials:
a time-dependent scale factor $R(t)=t^p$, mimicking
expansion or collapse; 
and strictly periodic driving, a crude model of the oscillations
that accompany the final stages of relaxation.
By computing the values of short-time Liapunov exponents,
Henry and co-workers found that trajectories in the expanding/contracting
potential could mimic regular orbits
part of the time and stochastic orbits at other times.
Contraction tended
to make the effects of chaos stronger, and expansion 
tended to make the chaos weaker.
In the oscillatory model, orbits appeared to remain either
regular or chaotic for all times, although the periodicity
in the global potential seemed to induce some orbits to
become what Henry termed ``wildy chaotic,'' exhibiting
substantial changes in energy as they chaotically diffused.

Having demonstrated the importance of chaotic mixing
in stationary, weakly time dependent, and strongly
time dependent potentials, 
it remained only for Henry only to make explicit the
link between chaotic mixing and ``violent relaxation.''
Henry did not get quite this far before his death, but
he had a clear idea of how to proceed.
In this passage, he describes a simple and beautiful
scheme for establishing the connection:

{\narrower\smallskip\noindent
But what, if anything, might these conclusions imply about
violent relaxation?
At least crudely, one can visualize an evolution described
by the collisinless Boltzmann equation as involving a
collection of characteristics corresponding to  orbits
evolved in a specified time-dependent potential, ignoring
the fact that that potential is generated self consistently
$\ldots$ to 
the extent that this picture is valid, one might then
anticipate that the efficacy with which an initial ensemble
appproaches an equilibrium or near-equilibrium $\ldots$ will depend
on the degree to which the flow in the specified potential
is chaotic.
In particular, to the extent that the flow is chaotic,
one would expect a rapid and efficient approach towards
a near-equilibrium.\smallskip}

\noindent 
In Henry's scheme,
one would first carry out a fully self-consistent simulation
of collapse and virialization, via an $N$-body code say,
recording the gravitational potential on a grid in 
both space and time.
Returning to the initial conditions, one would then
select out initially localized ensembles of phase-space
points and evolve them forward in the previously-recorded
potential, this time ignoring the self-gravity of the ensemble.
The sum total of all such integrations would be a reproduction
of the self-consistent collapse, and by analyzing the rate of
approach of each ensemble to its near-invariant distribution, 
one could generate a potentially complete picture of
the way in which the properties of the phase-space flow
were related to the rate of ``violent relaxation.''
Shortly before his death, Henry told me that he was
hoping to carry out this program in collaboration with
a student, but apparently this wish never came to fruition.

Like most dynamicists, Henry was fascinated by the
concept of entropy.
Henry's ideas about entropy were complex, and I'm not sure
that I understood them completely, but overall I felt
that Henry was skeptical about the relevance of entropy
arguments to galactic dynamics.
Here I can not resist quoting from V. A. Antonov [22], 
whose skeptical view of entropy was similar, I think, to Henry's:

{\narrower\smallskip\noindent
True diffusion is well described by differential 
equations. On the contrary, mixing is not represented
in terms of differential operations.
There is the phase density before the mixing,
and the phase density after the mixing, but it is difficult
to define when and where the transmutation occurs.
We could not work out general and utilizable equations
of the mixing.
\smallskip}

\noindent
Henry understood that chaotic mixing is
irreversible, in the sense that an infinitely precise
fine-tuning of the velocities would  be required in order to
undo its effects.
This is a sort of entropy increase, and it implies
an evolution toward a state whose properties are in
some ways predictable.
But like Antonov, Henry realized that it is difficult to 
establish very general rules that link the initial and
final states of a stellar system that evolves
via collisionless relaxation,
in particular, rules that would allow one to make statements about
which final states are preferred.
Henry was critical [23], for instance, about a purported demonstration 
[24] of an ``$H$-theorem'' for collisionless systems:

{\narrower\smallskip\noindent
It is, moreover, clear physically that there exist `reasonable'
choices of initial data, such as those leading to nearly
homologous collapse, which exhibit nearly periodic motion; 
and for such data, one might anticipate that, after one
approximate period, $H$ will have returned very nearly
to its initial value $\ldots$ Because the $H$-functions
$\ldots$ need not increase
monotonically, they cannot be used to provide a useful characterization
of the continuous dynamics.
\smallskip }

Henry also made fundamental contributions 
to our understanding of another sort
of chaos characteristic of stellar systems.
Already in 1964 [25], Richard Miller had shown that the
trajectories of stars in small-$N$-body integrations
were generically chaotic, in the sense that the
$6N$-dimensional phase path was exponentially unstable to
small changes in the initial conditions.
In a series of papers [26-29],
Henry and collaborators carried out a systematic
numerical study of this instability.
They found that the time scale for growth of perturbations
tended to {\it decrease} with increasing $N$,
remaining of order the crossing time for 
values of $N$ as large as $4000$.
This result was consistent with Henry's earlier theoretical
arguments [30-32] and in contradiction
with a prediction of Gurzadyan \& Savvidy [33] that
the growth rate should fall as $1/N^{1/3}$.
(Henry's prediction that the growth rate should
increase with increasing $N$ has recently been verified
for values of $N$ as large as $10^5$ [34].)
Henry recognized that this generic instability of the $N$-body
equations did not necessarly imply that mixing or relaxation
would be efficient, since the exponential instability often
seemed to ``saturate'' on scales much smaller than the scale
of the system.
However he made the interesting point [7] that the existence
of the instability made it difficult to reconcile the 
$N$-body equations of motion with the collisionless
Boltzmann equation:

{\narrower\smallskip\noindent
This leads, however, to an important question of principle.
The $N$-body problem appears to be chaotic on a time scale
$\sim t_{cr}$ [the crossing time], but the flow associated
with the CBE is often integrable or near-integrable
in the sense that many or all of the characteristics
are regular, i.e., nonchaotic.
So what do the (often near-integrable) CBE characteristics
have to do with the true (chaotic) $N$-body problem?
\smallskip}

\noindent
Henry asked: In what sense do the $N$-body
equations of motion ``go over'' to the collisionless
Boltzmann equation as $N\rightarrow\infty$?
Henry considered several possible ways in which this might happen,
and concluded

{\narrower\smallskip\noindent
Given the fact that the $N$-body problem is chaotic on
a time scale $\sim t_{cr}$, it would seem reasonable to
conjecture that the orbits generated in two
different $N$-body realizations will diverge exponentially on a
time scale $\sim t_{cr}$ $\ldots$ However, one might nevertheless
expect that, for sufficiently large $N$, the ensemble average of the
different $N$-body orbits generated from the same $(x_0,v_0$) will 
closely track the CBE characteristic for some finite time.
In particular, one might conjecture that the rms configuration
space deviation between the $N$-body orbits and the CBE
characteristics will scale as
$$
\delta r_{rms}(t) \approx F(N)\exp(t/\tau),
$$
where $\tau\approx t_{cr}$, roughly independent of the total
particle number $N$, and where the prefactor $F(N)\rightarrow 0$ 
for $N\rightarrow\infty$.
\smallskip}

\noindent Henry was proposing here a ``weak'' correspondence
between the $N$-body and CBE descriptions, in the sense
that an ensemble average of the $N$-body trajectories
might mimic the orbit in the smoothed-out potential.
This suggestion was characteristically cautious,
and soon after, Henry and I. Sideris [35] numerically
demonstrated a stronger 
sense in which the $N$-body trajectories approach the
smooth-potential orbits:

{\narrower\smallskip\noindent
$\ldots$ there is a clear, quantifiable sense in which, as $N$ increases, 
chaotic orbits in the frozen-$N$ systems remain ``close to'' integrable 
characteristics in the smooth potential for progressively longer times. 
When viewed in configuration or velocity space, or as probed by 
collisionless invariants like angular momentum, frozen-$N$ orbits 
typically diverge from smooth potential characteristics as a power 
law in time, rather than exponentially, on a time scale $\sim N^pt_D$, 
with $p\sim 1/2$ and $t_D$ a characteristic dynamical, or crossing, time. 
\smallskip}

\noindent By ``frozen-$N$'' orbits, Henry meant trajectories
computed in a potential where the smooth density had been replaced
by a matrix of fixed point masses, the gravitational analog of
the Lorentz gas [36].
For large $N$, these experiments showed that almost all
trajectories, and not just their ensemble averages, tended
in their finite behavior toward the behavior of regular orbits, 
even though the growth of infinitesimal perturbations (as measured
by Liapunov exponents) remained large.
In one of his last papers [37], Henry and I. Sideris showed that 
the chaos associated with a smooth potential could often
be distinguished from the $N$-body chaos, in spite of the latter
having a much shorter growth time,
by comparing initially nearby orbits in 
a single $N$-body system, or by tracking orbits with the same initial 
conditions evolved in two different $N$-body realizations of the 
same smooth density.

\acknowl 
I thank Richard Miller and Haywood Smith for their careful reading 
of the manuscript.

\references

\ 1. Kandrup, H. E. 1989. Stochastic problems in stellar dynamics.
     PhD Thesis, University of Chicago.

\ 2. Kandrup, H. E. 2003. Chaos and chaotic phase mixing in galaxy 
     evolution and charged particle beams. 
     {\it In} Galaxies and Chaos. G. Contopoulos and N. Voglis, 
     Eds. Lecture Notes in Physics {\bf 626}: 154-168.

\ 3. Zwicky, F. 1939. On the formation of clusters of nebulae
     and the cosmological time scale. Proc. Natl. Acad. Sci.
     {\bf 25}: 604-609.

\ 4. King, I. R. 1962. The structure of star clusters. I. An 
     empirical density law. Astron. J. {\bf 67}: 471-485.

\ 5. H\'enon, M. 1964. L'Evolution initiale d'un amas sph\'erique.
    Ann. d'Astro- phys. {\bf 27}: 83-91.

\ 6. Lynden-Bell, D. 1967. Statistical mechanics of violent 
     relaxation in stellar systems. Mon. Not. R. Astron. Soc.
     {\bf 136}: 101-121.

\ 7. Kandrup, H. E. 1998. Collisionless relaxation in galactic
     dynamics and the evolution of long-range order.
     {\it In} Long-Range Correlations in Astrophysical Systems.
     J. R. Buchler, J. W. Dufty \& H. E. Kandrup, Eds.
     Ann. N. Y. Acad. Sci. {\bf 848}: 28-47.

\ 8. Kandrup, H. E. \& Mahon, M. E. 1994.
     Relaxation and stochasticity in a truncated Toda lattice.
     Phys. Rev. E {\bf 49}: 3735-3747.

\ 9. Kandrup, H. E. \& Mahon, M. E. 1994.
     Short times characterisations of stochasticity in nonintegrable 
     galactic potentials.
     Astron. Astrophys. {\bf 290}: 762-770.

\ 10. Mahon, M. E., Abernathy, R. A., Bradley, B. O. \& Kandrup, H. E. 1995.
      Transient ensemble dynamics in time-independent galactic potentials.
      Mon. Not. R. Astron. Soc. {\bf 275}: 443-453.

\ 11. Kandrup, H. E. \& Siopis, C. 2003.
      Chaos and chaotic phase mixing in cuspy triaxial potentials.
      Mon. Not. R. Astron. Soc. {\bf 345}: 727-742.  

\ 12. Kandrup, H. E. 1998.
      Phase mixing in time-independent Hamiltonian systems.
      Mon. Not. R. Astron. Soc. {\bf 301}: 960-974.

\ 13. Kandrup, H. E. 1999.
      Invariant distributions and collisionless equilibria.
      Mon. Not. R. Astron. Soc. {\bf 299}: 1139-1145.

\ 14. Jeans, J. H. 1915.
      On the theory of star-streaming and the structure of the universe.
      Mon. Not. R. Astron. Soc. {\bf 76}: 70-84.

\ 15. Poon, M. Y. \& Merritt, D. 
      A self-consistent study of triaxial black hole nuclei.
      Astrophys. J. {\bf 606}: 774-787.

\ 16. Habib, S., Kandrup, H. E. \& Mahon, E. M. 1996.
      Chaos and noise in a truncated Toda potential.
      Phys. Rev. E {\bf 53}: 5473-5476.

\ 17. Habib, S., Kandrup, H. E. \& Mahon, E. M. 1997.
      Chaos and Noise in Galactic Potentials.
      Astrophys. J. {\bf 480}: 155-166. 

\ 18. Kandrup, H. E., Pogorelov, I. V. \& Sideris, I. V. 2000.
      Chaotic mixing in noisy Hamiltonian systems.
      Mon. Not. R. Astron. Soc. {\bf 311}: 719-732.

\ 19. Kandrup, H. E. \& Drury, J. 1998.
      Chaos in cosmological Hamiltonians.
      Ann. N. Y. Acad. Sci. {\bf 867}: 306-000.

\ 20. Kandrup, H. E., Vass, I. M. \& Sideris, I. V. 2003,
      Mon. Not. R. Astron. Soc. {\bf 341}: 927-936.

\ 21. Terzi\'c, B. \& Kandrup, H. E. 2004.
      Orbital structure in oscillating galactic potentials.
      Mon. Not. R. Astron. Soc. {\bf 347}: 957-967.

\ 22. Antonov, V. A. 2000.
      Brouwer Award Lecture (unpublished).

\ 23. Kandrup, H. E. 1987,
      An $H$-theorem for violent relaxation?
      Mon. Not. R. Astron. Soc. {\bf 225}: 995-998.

\ 24. Tremaine, S., H\'enon, M. \& Lynden-Bell, D. 1996,
      $H$-functions and mixing in violent relaxation.
      Mon. Not. R. Astron. Soc. {\bf 219}: 285-297.

\ 25. Miller, R. H. 1964.
      Irreversibility in small stellar dynamical systems.      
      Astrophys. J. {\bf 140}, 250-256.

\ 26. Kandrup, H. E. \& Smith, H. 1991.
      On the sensitivity of the $N$-body problem to small changes
      in initial conditions.
      Astrophys. J. {\bf 374}: 255-265.

\ 27. Kandrup, H. E. \& Smith, H. 1992,
      On the sensitivity of the $N$-body problem to small changes
      in initial conditions. II.
      Astrophys. J. {\bf 386}: 635-645.

\ 28. Kandrup, H. E., Smith, H. \& Willmes, D. E. 1992,
      On the sensitivity of the $N$-body problem to small changes
      in initial conditions. III.
      Astrophys. J. {\bf 399}: 627-633.

\ 29. Kandrup, H. E., Mahon, M. E. \& Smith, H. 1994,
      On the sensitivity of the $N$-body problem to small changes
      in initial conditions. IV.
      Astrophys. J. {\bf 428}: 458-465.

\ 30. Kandrup, H. E. 1989.
      The time scale for ``mixing" in a stellar dynamical system.
      Phys. Lett. A. {\bf 140}: 97-100.

\ 31. Kandrup, H. E. 1990.
      How fast can a galaxy ``mix"?
      Physica A {\bf 169}: 73-94.

\ 32. Kandrup, H. E. 1990.
      Divergence of nearby trajectories for the gravitational N-body problem.
      Astrophys. J. {\bf 364}: 420-425.

\ 33. Gurzadian, V. G. \& Savvidy, G. K. 1986.
      Collective relaxation of stellar systems.
      Astron. Astrophys. {\bf 160}: 203-210.

\ 34. Hemsendorf, M. \& Merritt, D. 2002
      Instability of the gravitational $N$-body problem in the
      large-$N$ limit.
      Astrohys. J. {\bf 580}: 606-609.

\ 35. Kandrup, H. E. \& Sideris, I. V. 2001.
      Chaos and the continuum limit in the gravitational $N$-body
      problem: Integrable potentials.
      Phys. Rev. E. {\bf 64}: 056209-1--11.

\ 36. Lorentz, H. A. 1905. The motion of electrons in metallic
     bodies. Proc. Amst. Acad. {\bf 7}: 438-453. 

\ 37. Kandrup, H. E. \& Sideris, I. V. 2003.
      Smooth potential chaos and $N$-body simulations.
      Astrophys. J. {\bf 585}: 244-249.

\bye